\documentclass[page-classic,a4paper]{myclass}

\title{Quantitative assessment of non-conservative radiation forces in
an optical trap} \shorttitle{Quantitative assessment of
non-conservative radiation forces in an optical trap}

\author{Giuseppe Pesce\inst{1,3} \and Giorgio Volpe\inst{2} \and Anna
Chiara De Luca\inst{1,3} \and Giulia Rusciano\inst{1,3} \and Giovanni
Volpe\inst{4,5}}

\shortauthor{G. Pesce et al.}

\institute{ \inst{1} Dipartimento di Scienze Fisiche, Universit\`{a}
di Napoli ``Federico II", Complesso Universitario Monte S. Angelo, Via
Cintia, 80126 Napoli, Italy\\ \inst{2} CNISM - Consorzio Nazionale
Interuniversitario per le Scienze Fisiche della Materia - Sede di
Napoli \\ \inst{3} ICFO - Institut de Ciencies Fotoniques,
Mediterranean Technology Park, 08860, \\Castelldefels (Barcelona),
Spain\\ \inst{4} Max-Planck-Institut f\"{u}r Metallforschung,
Heisenbergstr. 3, 70569 Stuttgart, Germany \\ \inst{5} 2. Physikalisches Institut, Universität Stuttgart, Pfaffenwaldring 57, 70550 Stuttgart, Germany } 


\abstract{The forces acting on an optically trapped particle are
usually assumed to be conservative.  However, the presence of a
non-conservative component has recently been demonstrated.  Here we
propose a technique that permits one to quantify the contribution of
such a non-conservative component.  This is an extension of a standard
calibration technique for optical tweezers and, therefore, can easily
become a standard test to verify the conservative optical force
assumption. Using this technique we have analyzed optically trapped
particles of different size under different trapping conditions. We
conclude that the non-conservative effects are effectively negligible
and do not affect the standard calibration procedure, unless for
extremely low-power trapping, far away from the trapping regimes
usually used in experiments.}

\begin{document}

\maketitle

\section{Introduction}

The detection and measurement of forces and torques in microscopic
systems is an important goal in many areas such as biophysics,
colloidal physics and hydrodynamics of small systems.  Since 1993, the
photonic force microscope (PFM) has become a standard tool to probe
such forces \cite{Ghislain1993,Ghislain1994,Neumann2004}. A typical
PFM setup comprises an optical trap -- an highly-focused
Gaussian light beam -- that holds a probe -- a dielectric or metallic
particle of micrometer size -- and a position sensing system.  Using a
PFM it has been possible to measure forces as small as $\mathrm{25\,
fN}$ \cite{Rohrbach2005} and torques as small as { $\mathrm{4000 \, fN
\cdot nm}$ \cite{Volpe2006}.}

\begin{figure} \onefigure[width=0.9\linewidth]{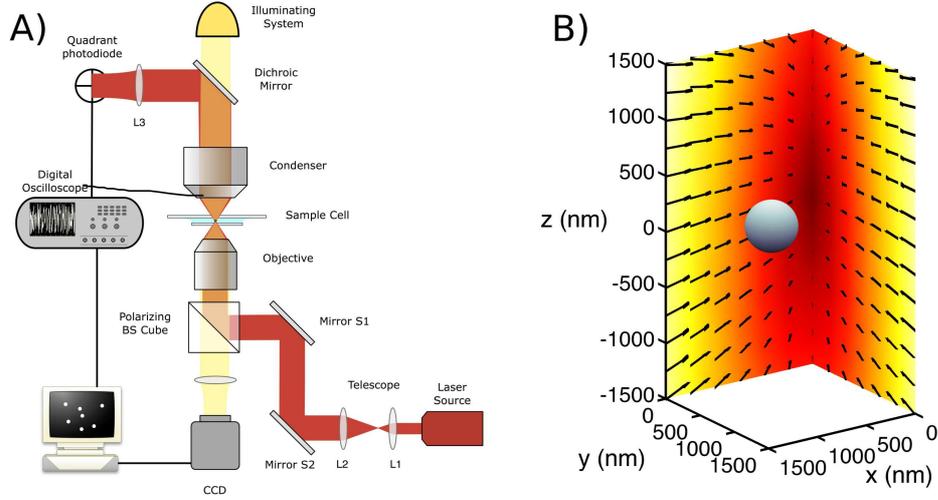}
\caption{(A) Schematics of the experimental setup. (B) Forces acting
on a colloidal particle held by an optical trap. The lines represent
the force-field and the color the modulus of the force.  A
slight bending of the force lines due to the non-conservative
component of the force-field can be observed [see also Fig. 2].}
\label{fig.1}
\end{figure}

In order to assess the mechanical properties of microscopic systems,
the first step is always to have an accurately calibrated optical
probe.  Modelling the interaction between the light of a focused laser
beam and an extended dielectric or metallic object can be a
complicated task \cite{Mazolli2003}.  The electromagnetic theory is
relatively straightforward for the Rayleigh and the geometrical optics
regimes\cite{HaradaOptComm96,Ashkin1992}. However, most applications
of PFM involve particles whose characteristic size is comparable to
the wavelength of the light employed.  In this case the exact
solutions for the force-field are cumbersome to come by.  Fortunately
there are several straightforward methods to experimentally measure
the trap parameters -- i.e., the trap stiffness and the conversion
factor between voltage and length -- and, therefore, the force exerted
by the optical tweezers on an object. The most commonly employed
methods are the \emph{drag force method}, the \emph{equipartition
method}, the \emph{potential analysis method} and the \emph{power
spectrum or correlation method} \cite{Visscher1996}. The latter two
\cite{Berg-Sorensen2004,Volpe2007PRE}, in particular, are usually
considered the most reliable ones.

An implicit assumption of all these calibration methods is that, for
small displacements of the probe from the center of an optical trap,
the restoring force is proportional to the displacement.  Hence, an
optical trap is assumed to act on the probe like a Hookeian spring
with a fixed stiffness.  This condition implicates that the the
force-field produced by the optical forces must be conservative,
excluding the possibility of a rotational component.  This is actually
true to a great extend in the plane perpendicular to the beam
propagation direction (e.g., x-y plane in Fig. 1(B)) for a
standard optical trap generated by a Gaussian beam.  However, this
has been shown not to be true in a plane parallel to the beam
propagation \cite{Ashkin1992,Merenda2006,Roichman2008} (e.g., x-z
plane in Fig. 1(B)).

Back in 1992 Ashkin already pointed out that, in principle,
scattering forces in optical tweezers do not conserve mechanical
energy, and that this could have some measurable consequences
\cite{Ashkin1992}. In particular, this non-conservative force would
produce a dependence of the axial equilibrium position of a trapped
micro-sphere as a function of its transverse position in the trapping
beam (see Fig. 10(C) of Ref. \cite{Ashkin1992}); such prediction was
first confirmed by Merenda and colleagues \cite{Merenda2006}.
Recently Roichman and colleagues \cite{Roichman2008} have directly
investigated the non-conservative component and have discussed the
implications that this might have for optical tweezers-based
experiments making use of the thermal fluctuations in the calibration
procedure.

Here we propose a technique that permits one to evaluate the relative
weight of the non-conservative component of the optical forces. It is
based on a previous work \cite{Volpe2007PRE}, where it was proposed an
enhancement of the PFM to measure force-fields with a non-conservative
component. We use this technique to analyze various optically trapped
particles in different trapping conditions.  The main result is that
the non-conservative effects are effectively negligible and do not
affect the standard calibration procedure, unless for extremely
low-power trapping, far away from the trapping regimes usually used in
experiments.

\section{Theory}

Assuming a very low Reynolds number regime
\cite{Purcell1977,HappelBrenner}, the motion of a Brownian particle in
the presence of an optical force-field can be described by the
vectorial Langevin equation
\begin{equation}
\label{e:ForceFieldBM} \mathbf{\dot{r}}(t) = \frac{1}{\gamma}
\mathbf{f}(\mathbf{r}(t)) + \sqrt{2 D} \mathbf{h}(t),
\end{equation} where $\mathbf{r}(t)$ is the probe position and
$\mathbf{f}(\mathbf{r})$ is the optical force acting on the particle,
which depends on the position of the particle itself, of course since
$\mathbf{r}$ is time-dependent then also $\mathbf{f}$ varies over
time, $\gamma = 3\pi d \sigma$ is its friction coefficient, $d$ is its
diameter, $\sigma$ is the medium viscosity, $\sqrt{2 D} \gamma
\mathbf{h}(t)$ is a vector of independent white Gaussian random
processes describing the Brownian forces, $D = k_B T / \gamma$ is the
diffusion coefficient, $T$ is the absolute temperature, and $k_{B}$ is
the Boltzmann constant.

The system that we are going to characterize is radially symmetric. It
is, therefore, easier to study it in cylindrical coordinates.
Eq.(\ref{e:ForceFieldBM}) can be projected in cylindrical coordinates
as
\begin{equation} \left\{ {\begin{array}{*{20}c} {\mathop \rho
\limits^. (t) = \frac{1}{\gamma} f_{\rho}(\rho,\theta,z) + \sqrt {2D}
h_\rho (t)} \\ {\mathop \theta \limits^. (t) = \frac{1}{\gamma \rho
(t)} f_{\theta}(\rho,\theta,z) + \sqrt {2D} \frac{{h_\theta
(t)}}{{\rho (t)}}} \\ {\mathop z\limits^. (t) = \frac{1}{\gamma}
f_{z}(\rho,\theta,z) + \sqrt {2D} h_z (t)} \\
\end{array}} \right.
\end{equation} where $h_\rho (t)$, $h_\theta (t)$ and $h_z (t)$ are
independent white Gaussian random processes with unitary variance.

Since an optical trap generated by a Gaussian beam is symmetrical, the
particle is effectively diffusing freely with respect to the
coordinate $\theta$ and we can assume that $f_{\theta}(\rho,\theta,z)
= 0$. We can, therefore, study the movement of the particle only with
respect to the coordinates $r$ and $z$.  Following a procedure similar
to the one in Ref. \cite{Volpe2007PRE}, to which we refer for more
details, we can linearize the force-field near the equilibrium
position $(\rho_0, z_0) = (0,0)$ and rewrite the Brownian particle
equations of motion as
\begin{equation} \left\{ {\begin{array}{*{20}c} {\mathop \rho
\limits^. (t) = -\frac{k_\rho}{\gamma} \rho(t) + \epsilon
\frac{k_\rho}{\gamma} z(t) + \sqrt {2D} h_\rho (t)} \\ {\mathop
z\limits^. (t) = - \epsilon \frac{k_\rho}{\gamma} \rho(t) - \eta
\frac{k_\rho}{\gamma} z(t) + \sqrt {2D} h_z (t)} \\
\end{array}} \right.
\end{equation} where $k_\rho$ is the optical trap stiffness in the x-y
plane, $\eta k_\rho$ is the trap stiffness along the z axis, $\eta$ is
the ratio between the trap stiffness in the x-y plane and the one
along z, which is typically $\sim 0.1$, and $\epsilon$ represent the
relative contribution of the non-conservative component of the
force-field.

\begin{figure} \onefigure[width=0.9\linewidth]{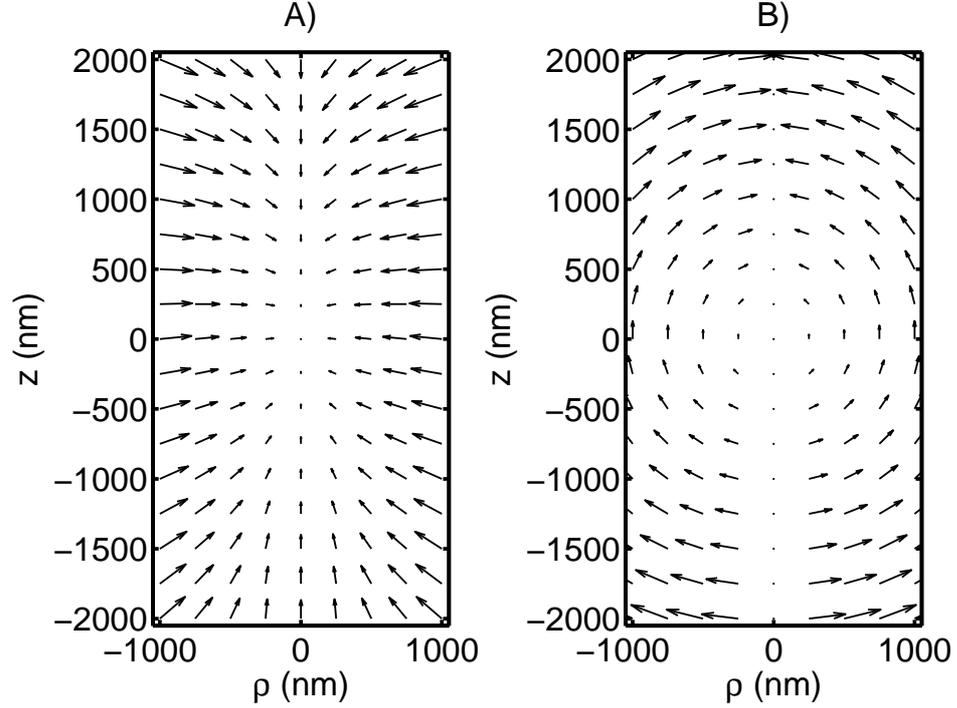}
\caption{(A) Force field generated by an optical trap in the
presence of a rotational component and (B) the rotational part of the
force-field with $\eta = 0.3$ and $\epsilon = 0.1$. Note that these
values are larger than typical ones, which are usually about $\eta
\sim 0.1$ and $\epsilon \sim 0.05$, in order to have a clearer
qualitative picture of the way they affect the force-field.}
\label{fig.2}
\end{figure}

The terms $-\frac{k_\rho}{\gamma} \rho(t)$ and $- \eta
\frac{k_\rho}{\gamma} z(t)$ represent the elastic restoring forces
which are the conservative part of the force-field and the terms $+
\epsilon \frac{k_\rho}{\gamma} z(t)$ and $- \epsilon
\frac{k_\rho}{\gamma} \rho(t)$ represent the non-conservative part of
the field.  In Fig. 2(A) an example of such a force-field is drawn.
In Fig. 2(B) only the non-conservative component is depicted.

The statistical analysis of the Brownian trajectories permits one to
reconstruct the force-field acting on the particle. Here we will use
the autocorrelation functions (ACFs) of the radial ($\rho$) and axial
($z$) particle position, and the difference between the two
cross-correlation functions (DCCF) between the the radial ($\rho$) and
axial ($z$) particle position.

Assuming $\epsilon << 1$ and $\epsilon << \eta$, the ACFs for $\rho$
and $z$ are decoupled
\begin{equation} \label{e:ACF_rho} \mathrm{ACF}_{\rho} (\tau ) =
\frac{\gamma D}{k_{\rho}}\exp{\left( -\frac{k_{\rho}}{\gamma}|\tau|
\right)}
\end{equation} and
\begin{equation} \label{e:ACF_z} \mathrm{ACF}_{z} (\tau ) =
\frac{\gamma D}{\eta k_{\rho}}\exp{\left( -\frac{\eta
k_{\rho}}{\gamma}|\tau| \right)}.
\end{equation} These expressions are the standard ones for a
conservative force-field and are independent from $\epsilon$
\cite{Volpe2007PRE}.  The DCCF is
\begin{eqnarray}\label{e:CCF} \mathrm{DCCF}_{\rho z} (\tau )
=4D\frac{{\varepsilon \gamma }}{{(1 + \eta ) k_\rho }}\exp \left( { -
\frac{{(1 + \eta )k_\rho }}{{2\gamma }}\left| \tau \right|}
\right)\cdot \nonumber \\ \cdot \frac{ \sinh{\left(
\frac{k_{\rho}}{2\gamma} \sqrt{|(1-\eta)^2 - 4\varepsilon^2|} \tau
\right)} } { \sqrt{|(1-\eta)^2 - 4\varepsilon^2|} }.
\end{eqnarray}

Equations (\ref{e:ACF_rho}), (\ref{e:ACF_z}) and (\ref{e:CCF}) can be
used to fit the values for $k_{\rho}$, $\eta$, and $\epsilon$, and
therefore to reconstruct the force-field up to the first order.  In
particular, the ACFs (\ref{e:ACF_rho}) and (\ref{e:ACF_z}) can be used
directly to fit the values of $k_{\rho}$ and $\eta$.  Once these
values are estimated, it is possible to fit the value $\epsilon$ by
using the slope of the DCCF (\ref{e:CCF}) around $\tau = 0$.

Once the force-field parameters have been fitted, it is possible to
calculate the torque acting on the particle due to the presence of the
rotational component of the force-field \cite{Volpe2006,Volpe2007PRE}
as
\begin{equation} \label{e:torque} T = \epsilon k_{\rho} \left[
\mathrm{Var(\rho) + Var(z)} \right],
\end{equation} and the circulation rate as
\begin{equation} \label{e:omega} \Omega = \epsilon
\frac{k_{\rho}}{2\pi \gamma}.
\end{equation}

\section{Experimental setup}

The experimental setup, shown in Fig. 1(A), is described in details in
Ref. \cite{PesceRSI05}.  The PFM comprises a home-made optical
microscope with a high-numerical-aperture water-immersion objective
lens (Olympus, UPLAPO60XW3, NA=1.2) and a frequency and amplitude
stabilized Nd-YAG laser ($\mathrm{\lambda=1.064\,\mu m}$,
$\mathrm{500\,mW}$ maximum output power, Innolight Mephisto).

Polystyrene micro-spheres (Serva Electrophoresis, $\mathrm{1.06
g/cm^3}$ density, 1.65 refractive index) with a diameter of
$\mathrm{0.45\pm0.01\,\mu m}$ and $\mathrm{1.25\pm0.05\, \mu m}$ were
diluted in distilled water to a final concentration of a few
particles/$\mathrm{\mu l}$. A droplet ($\mathrm{100\, \mu l}$) of such
solution was placed between a $150 \, \mu m$-thick coverslip and a
microscope slide, which were separated by a $100\, \mu m$-thick
parafilm spacer and sealed with vacuum grease to prevent evaporation
and contamination. Such sample cell was mounted on a closed-loop
piezoelectric stage (Physik Instrumente PI-517.3CL), which allowed
movements with nanometer resolution. The sample temperature was
continuously monitored using a calibrated NTC thermistor positioned on
the top surface of the microscope slide and remained constant within
0.2 degrees during each complete set of measurements.

A micro-sphere was trapped and positioned in the middle of the sample
cell, i.e. far away from the glass surfaces to avoid
hydrodynamic effects on the bead motion \cite{Berg-Sorensen2004}.
Its 3D position was monitored through the forward scattered light
imaged on a a InGaAs Quadrant Photodiode (QPD, Hamamatsu G6849) at the
back focal plane of the condenser lens \cite{GittesOL98}, using a
digital oscilloscope (Tektronix TDS5034B) for data-acquisition.  The
QPD-response was linear for displacements up to $\mathrm{300\, nm}$
($\mathrm{2\, nm}$ resolution, $\mathrm{250\, kHz}$ bandwidth). The
conversion factor from voltage to distance was calibrated using the
power spectral density method \cite{BuoscioloOC04}.  We
excluded significant deviations from a harmonic trapping profile by
verifying that the trapping potential could be fitted to such a
profile in all the three directions.

\section{Experimental results}

We performed the experiments using particles with diameter
$\mathrm{0.45\,\mu m}$ and $\mathrm{1.25\, \mu m}$. Similar particles
are commonly employed in experiments that use optical traps
\cite{Neumann2004}. The particle positions were acquired at
$\mathrm{2.5\,kHz}$, which is above the cutoff frequency of the
particle motion in the optical trap and below the QPD bandwidth.

For a dataset of 2N+1 particle position -- i.e. $x_n$, $y_n$ and $z_n$
for $n = -N, ..., -1, 0, 1, ..., N$ at times $t_n = 0.4 \cdot n \,
\mathrm{ms}$ -- the experimental ACFs and DCCF are
\begin{equation} \mathrm{ACF}_{\rho}^{(e)} (\tau = 0.4 \cdot m) =
\sum_{n = -N}^{N} \rho_{m-n} \rho_{n},
\end{equation}
\begin{equation} \mathrm{ACF}_{z}^{(e)} (\tau = 0.4 \cdot m) = \sum_{n
= -N}^{N} z_{m-n} z_{n},
\end{equation} and
\begin{equation} \mathrm{DCCF}_{\rho,z}^{(e)} (\tau = 0.4 \cdot m) =
\sum_{n = -N}^{N} \rho_{m-n} z_{n} - z_{m-n} \rho_{n},
\end{equation} where $\rho_n = \sqrt{x_n^2+y_n^2}$ and $z_n$ are the
time-series of the particle position in cylindrical coordinates.

For medium-high optical power at the sample (few tens of milliwatts)
the $\mathrm{DCCF}_{\rho,z}^{(e)} (\tau) \cong 0$ within the
experimental error. This is an evidence that the contribution of the
non-conservative force-field component is effectively negligible, or
at least undetectable for acquisition times up to several tens of
minutes.

In order to have a non-vanishing DCCF the optical power at the sample
needed to be reduced down to a few milliwatts. Furthermore, even under
such low power a clearly non-vanishing DCCF was only obtained when the
acquisition time was increased up to $\mathrm{400\,s}$ (i.e., $2N+1
\approx \mathrm{1\times10^6}$ particle positions). It is an
important remark the fact that, due to extremely low value of the
rotational contribution to the total force-field, it is not
immediately evident from the traces of the particle motion. Indeed the
particle undergoes a random movement in the $\rho-z$ plane, where it
is not possible to distinguish the presence of a rotational component
without the aid of a statistical analysis such as the one we propose
(see videos in the supplementary materials).

\begin{figure} \onefigure[width=0.9\linewidth]{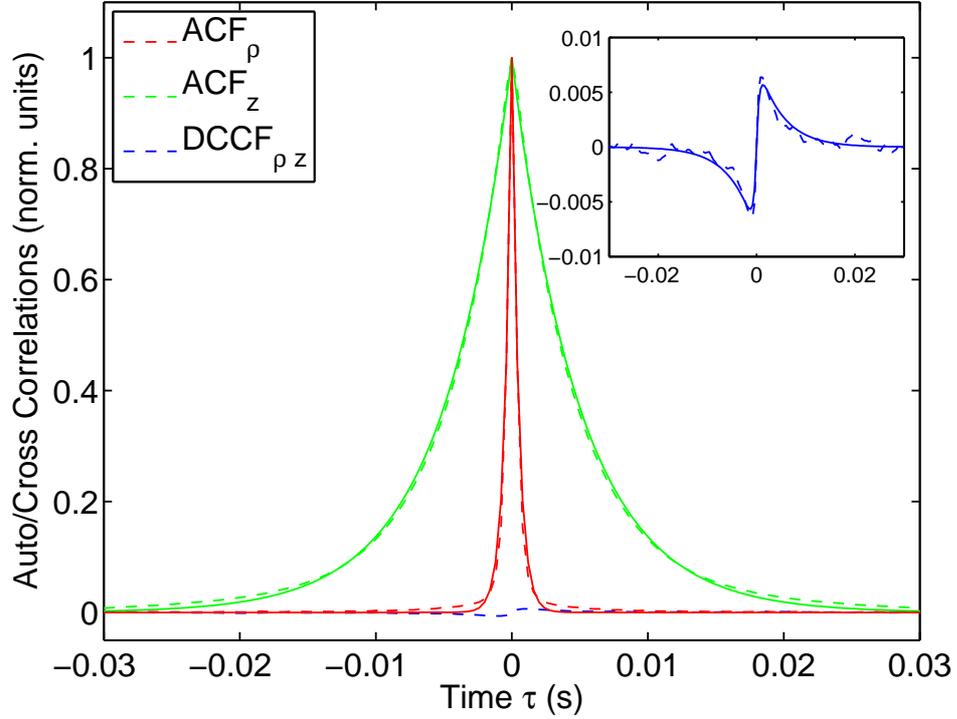}
\caption{Autocorrelation functions (ACFs) and difference of the
cross-correlation function (DCCF) for a $0.45\, \mu m$ colloidal
particle optically trapped with a laser power of 6 mW at sample (first
line of Tab. \ref{table}). Inset: close-up of the DCCF. Dashed lines
represent experimental data, while solid lines are the curves obtained
from the fit with Eq.(\ref{e:ACF_rho}), (\ref{e:ACF_z}) and
(\ref{e:CCF})}
\label{fig.3}
\end{figure}

In Fig. 3 the ACFs and DCCF are presented for a $\mathrm{0.45\,\mu m}$
diameter particle held in an optical trap with an optical power at the
sample of $\mathrm{6.0 \, mW}$ (average of 10
$\mathrm{400\,s}$-datasets). A good agreement is found between the
experimental and theoretical ACFs and DCCF. In particular, the
experimental and theoretical DCCF are very similar over all the range
of $\tau$ even though the fitting was performed only on the central
slope. The amplitude of the DCCF is very small compared to the ACF one
and the stiffness of the optical trap is quite low, only 5 $pN/\mu m$,
compared to common experiments with optical tweezers.

In a very weak optical trap the particle can explore regions far away
the trap-center. Thus the particle motion is more influenced by the
non-conservative force. This means that the less power is used the
more the DCCF amplitude is observed. This is shown in Figs. 4(A) where
we can see the behavior of the DCCF for a $\mathrm{0.45\,\mu m}$
diameter particle while decreasing the optical power.  The DCCF
amplitude increases as the power decreases. Furthermore, the range
over which the DCCF is not vanishing broadens. This
translates into an increase of the rotational component relative
weight for decreasing power as it is shown in Fig. 5(B).

\begin{figure} \onefigure[width=0.9\linewidth]{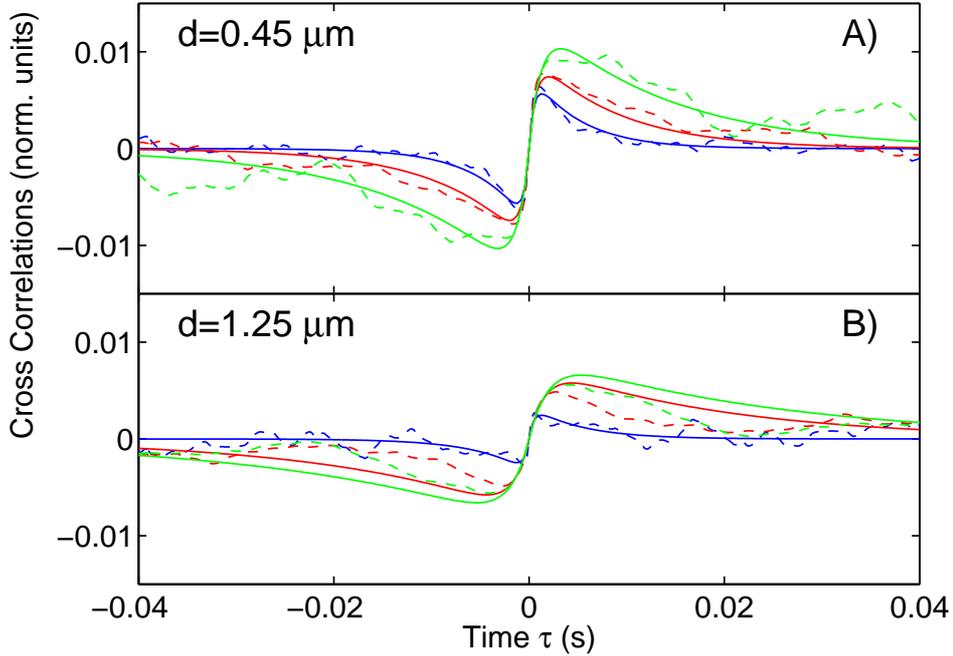}
\caption{Cross-correlation function for an optically trapped particle
whose diameter is (A) $0.45\, \mu m$ and (B) $1.25\, \mu m$. The
powers used are listed in Tab. \ref{table} and the color legend is:
P(blue) $>$ P(red) $>$ P(green). Dashed lines represent experimental
data, while solid lines are the curves obtaining from the fit}
\label{fig.4}
\end{figure}

In Fig. 4(B) we can see similar traces for a $\mathrm{1.25\, \mu
m}$. Again there is a good agreement between the experimental and
theoretical DCCF over a wide range, even though the fitting was
performed only on the central slope. Again the rotational
component relative weight increases for decreasing power (Fig. 5(B)).

It is worth to be noted that if we compare the DCCF curves for the two
diameters used at comparable stiffness, i.e. $k_{\rho}=2.9~ pN/\mu m$
for $d=0.45~\mu m$ and $k_{\rho}=3.6~ pN/\mu m$ for $d=1.25~\mu m$
(second and last line of Tab. \ref{table}) we can observe a slightly
larger amplitude for the smaller particle. Again this is a
confirmation that the effect is due to the larger volume explored by
the smaller particle with respect to the larger one.

In the Tab. \ref{table} the numerical values for the studied cases are
presented. The stiffness along the horizontal plane
$k_{\rho}$ is also shown in Fig. 5(A) where it can be appreciated the
fact that it is linear with respect to the optical power. The ratio
of the stiffness along $z$ and $\rho$ does not depend on the power
used, while a clear increase of the rotational component $\epsilon$ is
observed in both particle diameters used in this work.  We notice that
such values are similar to the ones reported in
Ref. \cite{Roichman2008}.  In their case the trap power is larger than
the one used in the present experiment, but for the different setup
the resulting stiffness is lower due to the fact that the particle
size is larger (2 $\mu m$).  Nevertheless the order of magnitude of
the circulation they observed is in agreement with our data.

In particular the torque associated to the non-conservative component
of the force-field is calculated according to Equation
\ref{e:torque}. The resulting values are extremely small 
(see Tab. \ref{table}), being actually orders of magnitude smaller 
than the ones previously reported, e.g., $4\times 10^{3} fN \cdot nm$ 
for the torque transfer
from a Laguerre-Gaussian beam to a Brownian particle \cite{Volpe2006},
$6 \times 10^3 fN \cdot nm$ for microscopic hydrodynamic flows
\cite{Volpe2008PRE}, $1 \times 10^4 fN \cdot nm$ for DNA twist
elasticity \cite{Bryant2003}, $5 \times 10^6 fN \cdot nm$ for the
movement of bacterial flagellar motors \cite{Berry1997}, $2\times 10^4
fN \cdot nm$ for the transfer of orbital optical angular momentum
\cite{VolkeSepulveda2002}, or $5\times 10^5 fN \cdot nm$ for the
transfer of spin optical angular momentum \cite{LaPorta2004}.
However, we must remark that in this letter we are focusing
on the non-conservative forces that arise in a standard optical trap
due to the fact that the trap is not perfectly harmonic in the
vertical plane, while some of the previously mentioned examples, such
as experiments with light beams that carry orbital or spin angular
momentum, refer to rather different situations in which the dominant
effect is the presence of a non-conservative (or rotational) force
field, which generates the particle movement.

\begin{largetable}
\caption{The experimental parameters and the values obtained from the
fit are reported here.}
\label{table}
\begin{center}
\begin{tabular}{ccccccc} \\ d & optical & $\rho$-stiffness &
$\mathrm{\frac{z-stiffness}{\rho-stiffness}}$ & rotational & torque &
circulation\\ ~ & power & ~ & ~ & component & ~ & rate \\ $d~(\mu m)$
& $P ~(mW) $& $k_\rho ~(pN/\mu m)$& $\eta$ & $\epsilon$ & $T ~
(fN\cdot nm)$ & $\Omega~ (Hz)$\\ \hline 0.45$\pm0.01$ & 6.0$\pm0.2$ &
5.0$\pm$0.1 & 16$\pm 1 \%$ & 1.0$\pm0.1~\%$ & 280$\pm$40 &
2.1$\pm$0.3\\ 0.45$\pm0.01$ & 3.2$\pm0.2$ & 2.9$\pm$0.1 & 15$\pm 1 \%$
& 1.7$\pm0.2~\%$ & 490$\pm$60 & 2.1$\pm$0.3\\ 0.45$\pm0.01$ &
1.0$\pm0.2$ & 1.68$\pm$0.07& 16$\pm 1\%$ & 2.4$\pm0.2~\%$ & 680$\pm$60
& 1.8$\pm$0.3\\ \hline 1.25$\pm0.05$ & 4.0$\pm0.2$ & 14.0$\pm$0.7&
14$\pm 1 \%$ & 0.7$\pm0.1~\%$ & 210$\pm$30 & 1.4$\pm$0.2\\
1.25$\pm0.05$ & 1.4$\pm0.2$ & 4.4$\pm$0.2 & 13$\pm 1 \%$ &
1.2$\pm0.1~\%$ & 420$\pm$70 & 0.8$\pm$0.1\\ 1.25$\pm0.05$ &
1.0$\pm0.2$ & 3.6$\pm$0.2 & 12$\pm 1 \%$ & 3.8$\pm0.3~\%$ &
1400$\pm$160 & 2.1$\pm$0.3
\end{tabular}
\end{center}
\end{largetable}

\begin{figure} \onefigure[width=0.9\linewidth]{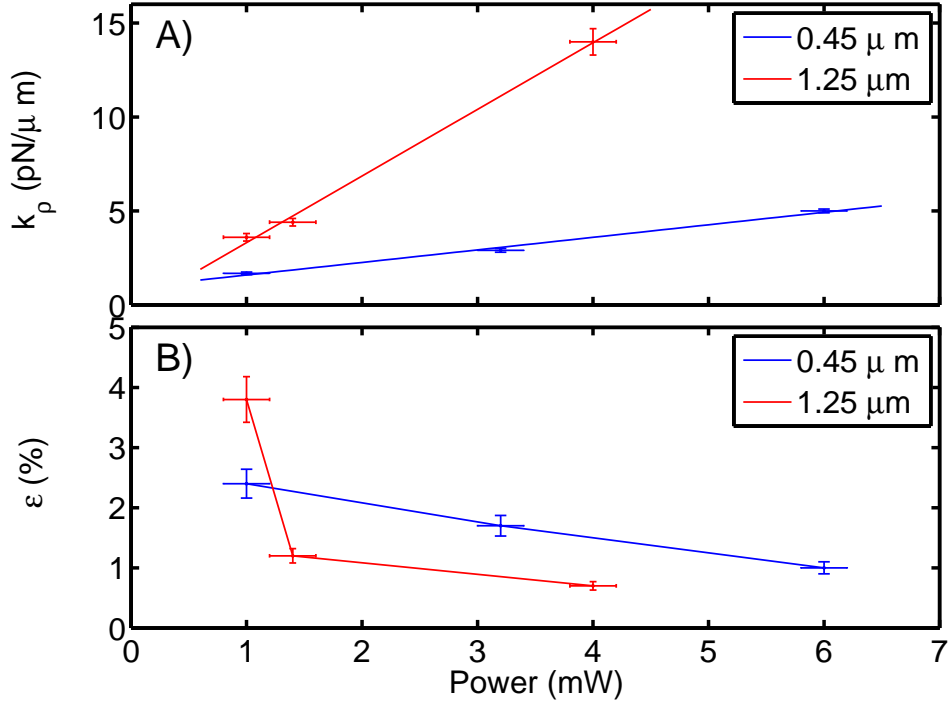}
\caption{(A) Behavior of the radial stiffness $k_{\rho}$ as function
of the laser power for the two kinds of particle used. The solid lines
are the linear fits. (B) Value of the relative contribution of the
non-conservative component $\epsilon$ as a function of the power
used. Notice that it increases for lower powers.}
\label{fig.5}
\end{figure}

\section{Conclusions}

The $\mathrm{DCCF}_{\rho,z}^{(e)} (\tau) \cong 0$ within the
experimental error is a clear evidence that the contribution of the
non-conservative force-field component is effectively negligible. This
is actually true in most practical experimental situations. Typical
optical tweezers experiments are performed with various tens of
milliwatts of optical power at the sample and acquiring data for at
most a few minutes \cite{Neumann2004}; however, we needed to decrease
the optical power at the sample down to a range of a few milliwatts
\emph{and} to increase the acquisition time up to several minutes in
order to see the signature of a non-conservative force-field component
in the DCCF.

In particular, for medium-high laser powers (from a few tens of
milliwatts at the sample onwards) the effect is undetectable even for
long acquisition times -- indeed, the value of $\epsilon$ steadily
decreases as the laser power is increased (see Tab. \ref{table} and
Fig. 5(B)).  This is due to the fact that the deviation from a
conservative force-field is larger far from the trap center, which is
explored more often in a weaker trap. Furthermore, this can also be
seen in Figs. 6 and 7 of Ref. \cite{Merenda2006}.

Therefore, we conclude that for the trapping regimes that are usually
employed in experiments, the effect of the non-conservative
force-field component are effectively negligible and do not affect the
standard calibration procedure.  Whenever a doubt is present, the
extension to the standard optical tweezers calibration techniques we
have proposed in this letter can be used to verify to what extend the
assumption of a conservative force-field is fulfilled.  In
particular this technique might become increasingly useful as optical
measurements of forces reach to ever smaller length and force scales,
because these nonequilibrium effects will become increasingly
noticeable.

\acknowledgments The authors acknowledge fruitful discussions with
Clemens Bechinger, Antonio Sasso and Dmitri Petrov. GR aknowledges
CNISM for her research fellowship.

\bibliographystyle{mybib}

\end{document}